\documentclass[reprint,showpacs,showkeys,amsmath,amssymb,superscriptaddress,aps,prl]{revtex4-1}

\usepackage{graphicx}
\usepackage{epstopdf}
\usepackage{color}

\begin{document}


\title{ Observation of Topological Hall Effect and Signature of Room Temperature Antiskyrmions in Mn-Ni-Ga D$_{2d}$ Heusler magnets}

\author{Subir Sen}
\thanks{Authors contributed equally}
\affiliation{School of Physical Sciences, National Institute of Science Education and Research, HBNI, Jatni-752050, India}
\author{Charanpreet Singh}
\thanks{Authors contributed equally}
\affiliation{School of Physical Sciences, National Institute of Science Education and Research, HBNI, Jatni-752050, India}
\author{Prashanta K. Mukharjee}
\affiliation{School of Physics, Indian Institute of Science Education and Research, Thiruvananthapuram, Kerala-695551, India}
\author{Ramesh Nath}
\affiliation{School of Physics, Indian Institute of Science Education and Research, Thiruvananthapuram, Kerala-695551, India}
\author{Ajaya K. Nayak}
\email{ajaya@niser.ac.in}
\affiliation{School of Physical Sciences, National Institute of Science Education and Research, HBNI, Jatni-752050, India}

\date{\today}

\begin{abstract}

Topologically stable nontrivial spin structures, such as skyrmions and antiskyrmions, display a large topological Hall effect owing to their quantized topological charge.  Here, we present the finding of a large topological Hall effect beyond room temperature in the tetragonal phase of a Mn-Ni-Ga based ferrimagnetic Heusler shape memory alloy system. The origin of the field induced topological phase, which is also evidenced by the appearance of dips in the ac-susceptibility measurements, is attributed to the presence of  magnetic antiskyrmions driven by $D_{2d}$ symmetry of the inverse Heusler tetragonal phase. Detailed micromagnetic simulations asserts that the antiskyrmionic phase is stabilized as a result of interplay among inhomogeneous Dzyaloshinskii-Moriya interaction, the Heisenberg exchange, and the magnetic anisotropy energy. The robustness of the present result is demonstrated by stabilizing the antiskyrmion hosting tetragonal phase up to a temperature as high as 550~K by marginally varying the chemical composition, thereby driving us a step closer to the realization of ferrimagnetic antiskyrmion based racetrack memory.   
\end{abstract}


\maketitle

In recent years, there is a significant interest towards non-collinear magnetism, where the local magnetic state can be {\it periodically} altered via spin transfer torque by passing a spin polarized current \cite{Parkin08}.  The prospect of non-collinear magnetism can be greatly enhanced when the aforementioned magnetic structure is topologically stable in nature. One of such spin textures is the recently discovered magnetic skyrmion, which is a vortex-like object with a swirling spin configuration \cite{Pfleiderer09, Yu10}. The topological nature of the skyrmions helps them to get decoupled from the crystal lattice, thereby assisting to move at much lower current density in comparison to that of domain walls \cite{Schulz12}. The topologically stable spin texture of the skyrmions is accompanied by a topological charge  $Q =\frac{1}{4\pi}\int {\bf m} \cdot (\frac{\partial {\bf m}}{\partial x} \times \frac{\partial {\bf m}}{\partial y}) dxdy= \pm 1$, where ${\bf m}$ is the unit vector along the local magnetization \cite{Nagaosa13}. When a conduction electron approaches a skyrmion, the spin of the electron tries to align with the local magnetization of the skyrmion owing to a large Hund's coupling. Consequently, the electrons experience a large fictitious magnetic field, resulting in an additional component to the observed Hall voltage, named as topological Hall effect (THE) \cite{Neubauer09}.  Effective fictitious magnetic field of 4000~T can be realized for a skyrmion of size 1~nm \cite{Kanazawa15}. Depending upon the topological charge of the skyrmion ($\pm1$), the topological Hall component adds or subtracts from the normal and anomalous Hall components to develop a hump or dip type of behavior in the total Hall voltage observed in various bulk materials and thin films \cite{Gallagher17,Matsuno16,Kanazawa15,Li13,Schulz12, Huang12,Kanazawa11,Neubauer09}. 

The topologically stable spin texture of the skyrmions arises from the spin-orbit interaction mediated Dzyaloshinskii-Moriya interaction (DMI), which competes with the Heisenberg exchange ($J$) and magnetic anisotropy to form a stable skyrmion lattice. The DMI energy that can be expressed as $\bf D_{ij} \cdot (\bf S_i \times \bf S_j)$, where $D$ is the DMI vector and $S_i$ and $S_j$ are spins at the $i^{th}$ and $j^{th}$ sites, respectively, exists in systems with broken inversion symmetry and large spin-orbit coupling \cite{Dzyaloshinskii57,Moriya60}. The magnetic materials with B20 and related crystal classes that possess intrinsic bulk DMI display Bloch-type skyrmions \cite{Pfleiderer09, Yu10, Yu11, Seki12, Tokunaga15}, whereas, most of the layered thin films with interfacial DMI and some bulk materials with suitable crystal symmetry host N\'{e}el skyrmions \cite{Heinze11, Sampaio13,Jiang15,Woo16,Soumyanarayanan17, Kezsmarki15,Kurumaji17}. Artificial magnetic skyrmions with nano-patterning is also realized without DMI \cite{Li14, Sun13, Miao14}.  A latest addition to the skyrmion family is the recently observed antiskyrmions in $D_{2d}$ crystal symmetry based inverse tetragonal Heusler compounds \cite{Nayak17}. The special crystal symmetry of these materials ensures an inhomogenious DMI vector ($D_x=-D_y$) in contrast to the homogeneous DMI observed in materials exhibiting Bloch and  N\'{e}el skyrmions ($D_x=D_y$) \cite{Leonov17,Hoffmann17,Camosi18}. 

It has been established that Mn$_2YZ$- based Heusler compounds display a non-centrosymmetric crystal structure \cite{Graf11}.  The DMI in these materials can be set up in case of Mn$_2YZ$ tetragonal Heusler compounds \cite{Meshcheriakova14,Nayak17}. The Heusler shape memory alloys (SMA) that undergo a martensite transtion from high temperature cubic to low temperature tetragonal phase, possess a great potential to host nontrivial spin texture like skyrmions. However, most of these alloys exhibit a modulated and centrosymmetric tetragonal structure that preclude DMI in the system \cite{Nayak14,Planes09, Yu15}. An asymmetric tetragonal structure with $D_{2d}$ symmetry can be stabilized in case of Mn$_2$NiGa when a single Mn atom in the Mn-Mn plane is replaced by Ni atom in Mn$_3$Ga, as shown in Fig. 1(a) \cite{Balke07,Nayak15, Liu05,Liu06}. The Mn sitting in Mn-Mn/Mn-Ni plane and the Mn at Mn-Ga plane align antiferromagnetically, account for the ferrimagnetic ordering in the system \cite{Barman08}. The existence of a more complex non-collinear spin structure and/or  the presence of slight antisite disorder intrinsic to most of the Heusler materials  can also result in the mismatch between experimentally observed moment with that of theoretical prediction \cite{Barman08}. In this letter we show that Mn-rich Mn-Ni-Ga based inverse Heusler system indeed displays a large topological Hall effect in the tetragonal phase, suggesting the presence of antiskyrmions in the system. 


\begin{figure} [tb!]
	\includegraphics[angle=0,width=8cm,clip]{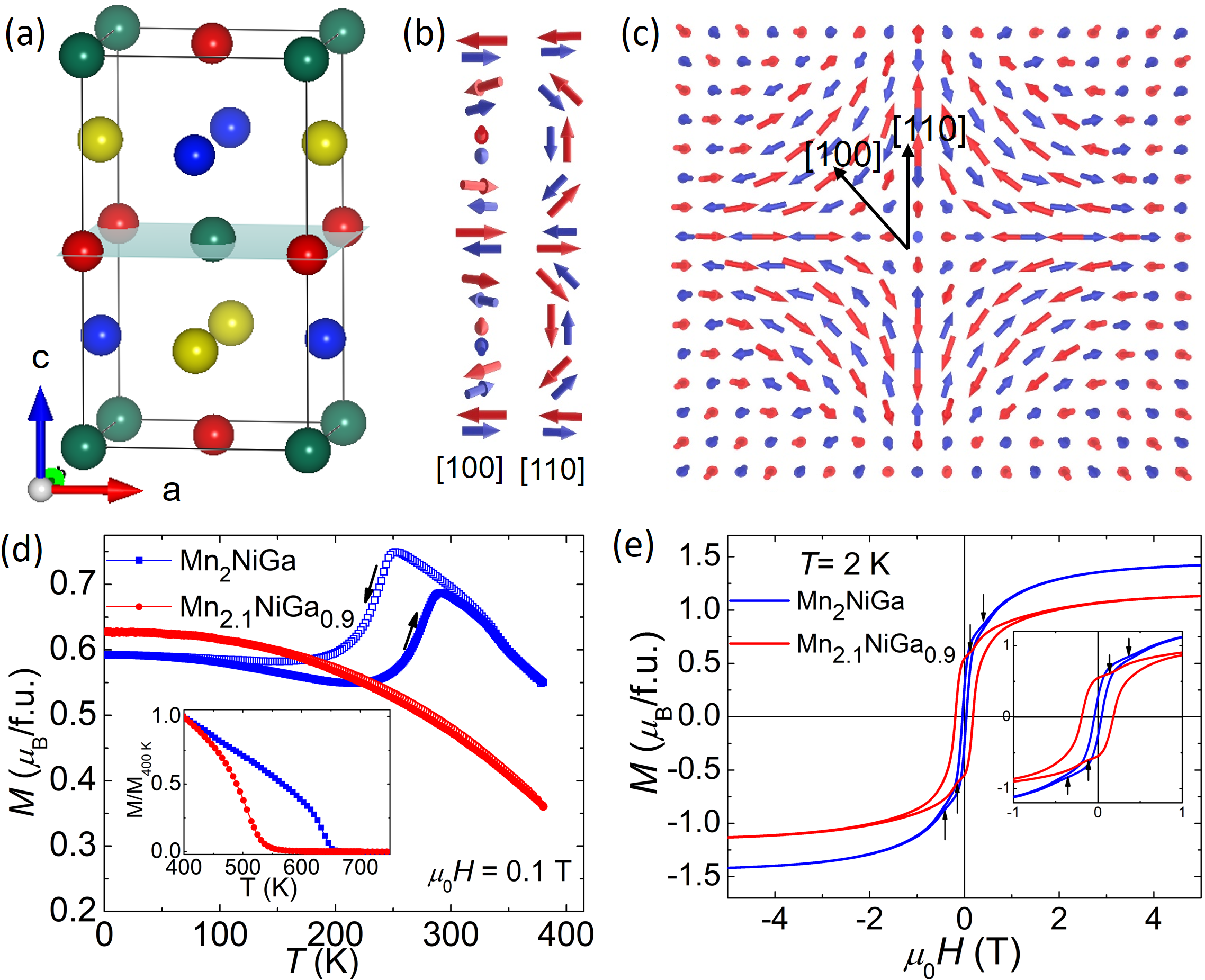}
	\caption{\label{FIG1}(Color online) (a) Crystal structure of Mn$_2$NiGa: MnI, Ga, MnII and Ni atoms are represented by red, green, blue and yellow balls, respectively. The $ab$-plane is shown in light green color. (b) Schematic representation of one-dimensional spin propagation along [100] (helix) and [110] (cycloid) in the $ab$-plane. (c) Spin configuration of a ferrimagnetic antiskyrmion in the present system. (d) Field cooled (FC) and field heating (FH) temperature dependence of magnetization [$M(T)$] for Mn$_2$NiGa and Mn$_{2.1}$NiGa$_{0.9}$ measured in 0.1~T field. Inset shows the high temperature  $M(T)$ curves showing the Curie temperature ($T_C$).  (d) Field dependent magnetization [$M(H)$] loops measured at 2~K for Mn$_2$NiGa and  Mn$_{2.1}$NiGa$_{0.9}$. The inset shows low field region of the data.}
\end{figure}

Like most of the Mn$_2$- based Heusler compounds, Mn$_2$NiGa exhibits a ferrimagnetic ordering with a Curie temperature of $\sim$ 650~K.  The  $D_{2d}$ symmetry of the tetragonal phase can ensure a competing iteraction between the DMI and the Heisenberg exchange that can result in an one-dimensional helix in the [100] direction and a cycloid along the [110] direction, as shown in Fig. 1(b). In this scenario, application of magnetic field can generate ferrimagnetic antiskyrmions as schematically demonstrsted in Fig. 1(c). The thermomagnetic $M(T)$ curves measured in field-cooled (FC) and field-heating (FH) modes up to 400~K for Mn$_2$NiGa and Mn$_{2.1}$NiGa$_{0.9}$ are depicted in Fig. 1 (d). The signature of structural transition in Mn$_2$NiGa can be seen from the presence of large hysteresis in cooling and heating $M(T)$ curves around 300~K, whereas, no such transition is found in Mn$_{2.1}$NiGa$_{0.9}$. Both the samples exhibit the  Curie temperature ($T_C$) well above the room temperature [inset of Fig. 1(d)].   An experimental signature that hints at the presence of an additional magnetic phase in both the samples was obtained from the isothermal magnetization $M(H)$ data which display a small kink marked by arrows in Fig. 1(e).  Since Mn$_2$NiGa transforms to the cubic phase above 300~K, this transition disappears for  $M(H)$ loops measured at higher temperatures. The magnetic moment of the present Mn$_2$NiGa matches well with the previous report \cite {Liu06}.


\begin{figure} [tb!]
	\includegraphics[angle=0,width=8cm,clip]{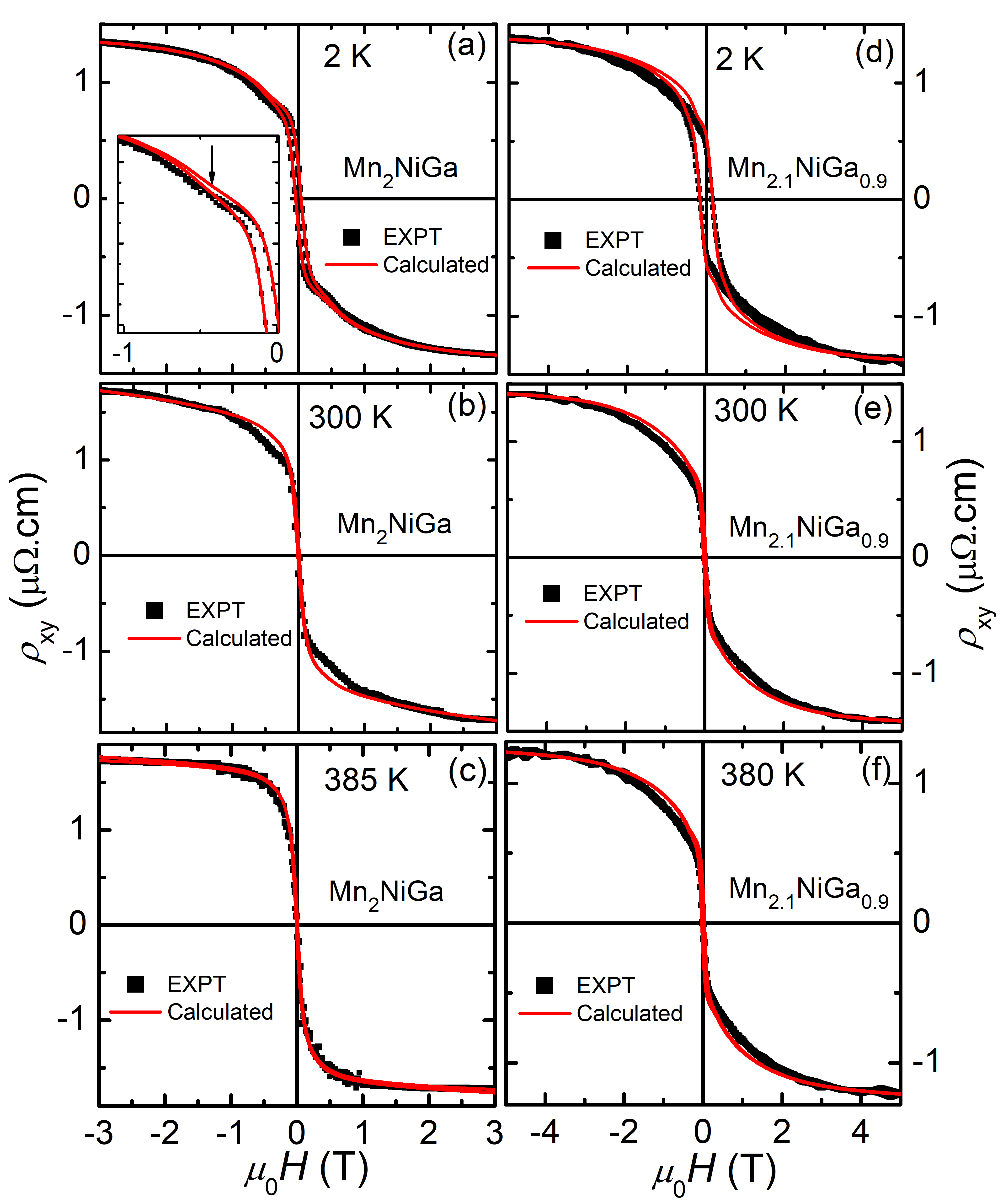}
	\caption{\label{FIG1}(Color online)  Field dependent Hall resistivity ($\rho_{xy}$, filled squares) measured at different temperatures for (a)-(c) Mn$_2$NiGa, (d)-(f) Mn$_{2.1}$NiGa$_{0.9}$. The inset of (a) shows an expanded view of the $\rho_{xy}$ in the second quadrant. The solid lines represent calculated Hall resistivity without topological Hall contribution as described in the text.}
\end{figure}

Motivated by the signature of magnetic phase transition in the $M(H)$ data, we have performed Hall effect measurements at different temperatures as depicted in Fig. 2. For Mn$_2$NiGa, the total Hall resistivity $\rho_{xy}$ exhibits a dip around 0.5~T for all temperatures $T \leq 300$~K [Fig. 2(a)\&(b)]. This peculiar behavior disappears for the $\rho_{xy}$ data collected above the martensite transition at $T=385$~K in the cubic phase [Fig. 2(c)]. Although the martensite transition sets the DMI in the tetragonal phase, it has no role in the observed anomaly in the $\rho_{xy}$ data. This is  demonstrated in another sample  Mn$_{2.1}$NiGa$_{0.9}$ that exhibits tetragonal phase in the whole temperature range up to the $T_{\text C}$, without any structural transition.  For this sample the  $\rho_{xy}$ data acquired up to 380~K, the highest possible measured temperature, display a similar dip kind of behavior around 1~T [Fig. 2(d)-(f)].

It is well known that the total Hall resistivity can be expressed as $\rho_{xy} = \rho_{N} + \rho_{AH} + \rho^T_{xy}$, where $\rho_{N}$, $\rho_{AH}$ and $\rho^T_{xy}$ are  normal, anomalous, and topological Hall resistivities, respectively. Normal Hall resistivity can be written as $\rho_{N} = R_{0}H$, where $R_0$ is the normal Hall coefficient. Anomalous Hall resistivity, which is in general directly proportional to the magnetization in a ferri-/ferromagnet, can be expressed in terms of the longitudinal resistivity ($\rho_{xx}$) and magnetization ($M$) as $\rho_{AH} = b\rho_{xx}^2 M$, where $b$ is a constant. The effect of skew scattering and side-jump on AHE in the present sample are not taken into consideration due to the fact that the longitudinal resistivity in the present bulk materials is too high and acan be neglected completely at high temperatures. In case of Mn$_2$NiGa,  the anomaly in the $\rho_{xy}$ data is only found for fields less than 1~T, whereas, Mn$_{2.1}$NiGa$_{0.9}$ displays such behavior for fields up to 2~T . Hence it is assumed that the high field $\rho_{xy}$ data do not consist of any $\rho^T_{xy}$ component. At high fields, $\rho_{xy}$ can be further simplified to  $\rho_{xy} = R_{0}H + b\rho_{xx}^2 M$. The linear fit between $\frac{\rho_{xy}}{H}$ and $\frac{\rho_{xx}^2 M}{H}$ gives us slope $b$ and intercept $R_0$. In the present case, the values of \textit{b} and $R_0$ are calculated by using  $\rho_{xy}$ data for $\mu_0H > \pm~3~T$. Afterwards, $\rho_{xy}$ sans $\rho^T_{xy}$  was calculated using $\rho_{xy} = R_{0}H + b\rho_{xx}^2 M$, as shown by red lines on the top of the experimental $\rho_{xy}$ curves in Fig. 2. It can be clearly seen that the experimental and the calculated $\rho_{xy}$ curves display a substantial difference at the field where both the magnetization and Hall resistivity exhibit dips, whereas, perfect matching is obtained for higher field regions. The calculated $\rho_{xy}$ was subtracted from the experimental $\rho_{xy}$ to obtain $\rho^T_{xy}$ as plotted in Fig. 4(a)\&(b). The validity of the extraction of THE by the present method is well verified in case of Mn$_{1.8}$Ni$_{1.2}$Ga, where both the experimental and calculated curves match at all field regime, as this sample does not exhibit any anomaly in Hall effect measurements.

\begin{figure} [tb!]
	\includegraphics[angle=0,width=8cm,clip]{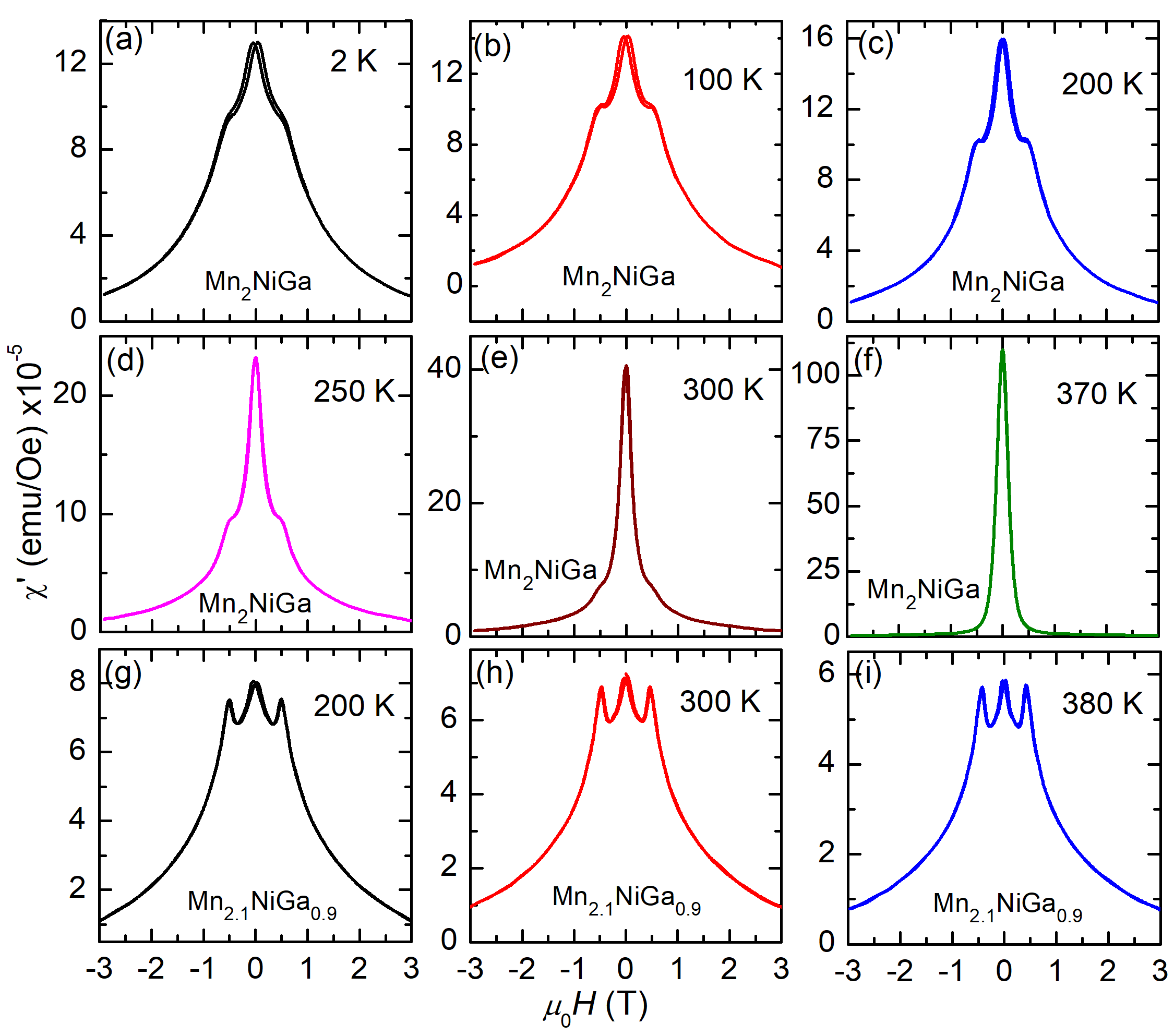}
	\caption{\label{FIG1}(Color online)  Field dependent ac-susceptibility measured at different temperatures for; (a)-(f) Mn$_{2}$NiGa and  (g)-(i) Mn$_{2.1}$NiGa$_{0.9}$. }
\end{figure}


For a deep understanding of the origin of the observed topological Hall effect, we have carried out ac-susceptibility measurements  which  have been extensively used to characterize skyrmions in several materials \cite{Kurumaji17,Wilhelm11,Bauer12}. In the present case, the real part of the ac-susceptibility, $\chi^{'} (H)$, for Mn$_2$NiGa exhibits a dip/peak type behavior around the fields where a large topological Hall effect is found. The magnitude of this dip/peak behavior initially increases with increasing temperature before getting slowly suppressed for $T\geq 250$~K due to the presence of a small amount of cubic phase with higher magnetic suceptibility.   The dip/peak completely vanishes at 370~K  in the cubic phase.  For Mn$_{2.1}$NiGa$_{0.9}$, a pronounced and well defined dip/peak can be found up to 380~K, suggesting the presence of magnetic antiskyrmions in the present system. It is worth to mention that the tetragonal Heusler compound Mn$_{1.4}$Pt$_{0.9}$Pd$_{0.1}$Sn that displays antiskyrmions up to room temperature~\cite{Nayak17} also exhibits a similar  behavior in the ac-susceptibility data \cite{Manna18}. The robostness  of the ac-susceptibility measurements is vindicated in case of Mn$_{1.8}$Ni$_{1.2}$Ga that does not exhibit any anomaly in $\chi^{'} (H)$ as no THE is found in this sample.


The occurrence of large topological Hall effect that is underpinned by the observation of dips/peaks in the ac-susceptibility data in the present material lends firm support for the existence of some non-trivial spin texture, such as, skyrmions. The $D_{2d}$ crystal symmetry of the present system ensures an anisotropic DMI with $D_x=-D_y$, thereby leading to antiskyrmions. In order to gain more insights into the magnetic field and temperature dependence of the antiskyrmions in the present system, we have plotted $\rho^T_{xy}$ at different magnetic fields as shown in Fig. 4(a)\&(b). In general, the topological Hall effect scales (i) directly with the density and (ii) inversely  with the size of the skyrmions/antiskyrmions. Since for a given system the size of the skyrmions/antiskyrmions remains almost constant, the enhanced $\rho^T_{xy}$ at room temperatures for  Mn$_2$NiGa can be attributed to a significant increase in antiskyrmion density due to the higher nucleation ability of antiskyrmions at the tetragonal to cubic phase transition. This can be understood from the fact that in case of a bulk system the nucleation probability of skyrmions/antiskyrmions increases around the magnetic ordering temperature \cite{Nayak17,Pfleiderer09,Yu11,Tokunaga15}. In case of Mn$_{2.1}$NiGa$_{0.9}$, the magnitude of $\rho^T_{xy}$ almost remains constant up to the highest measured temperature of 380~K.
Figure 4(c)\&(d) represent the $H-T$ phase diagrams for Mn$_2$NiGa and Mn$_{2.1}$NiGa$_{0.9}$. For Mn$_2$NiGa the THE almost vanishes for a field of about 1.5~T, whereas, Mn$_{2.1}$NiGa$_{0.9}$ exhibits THE  for field as high as 3~T.

\begin{figure} [tb!]
	\includegraphics[angle=0,width=8cm,clip]{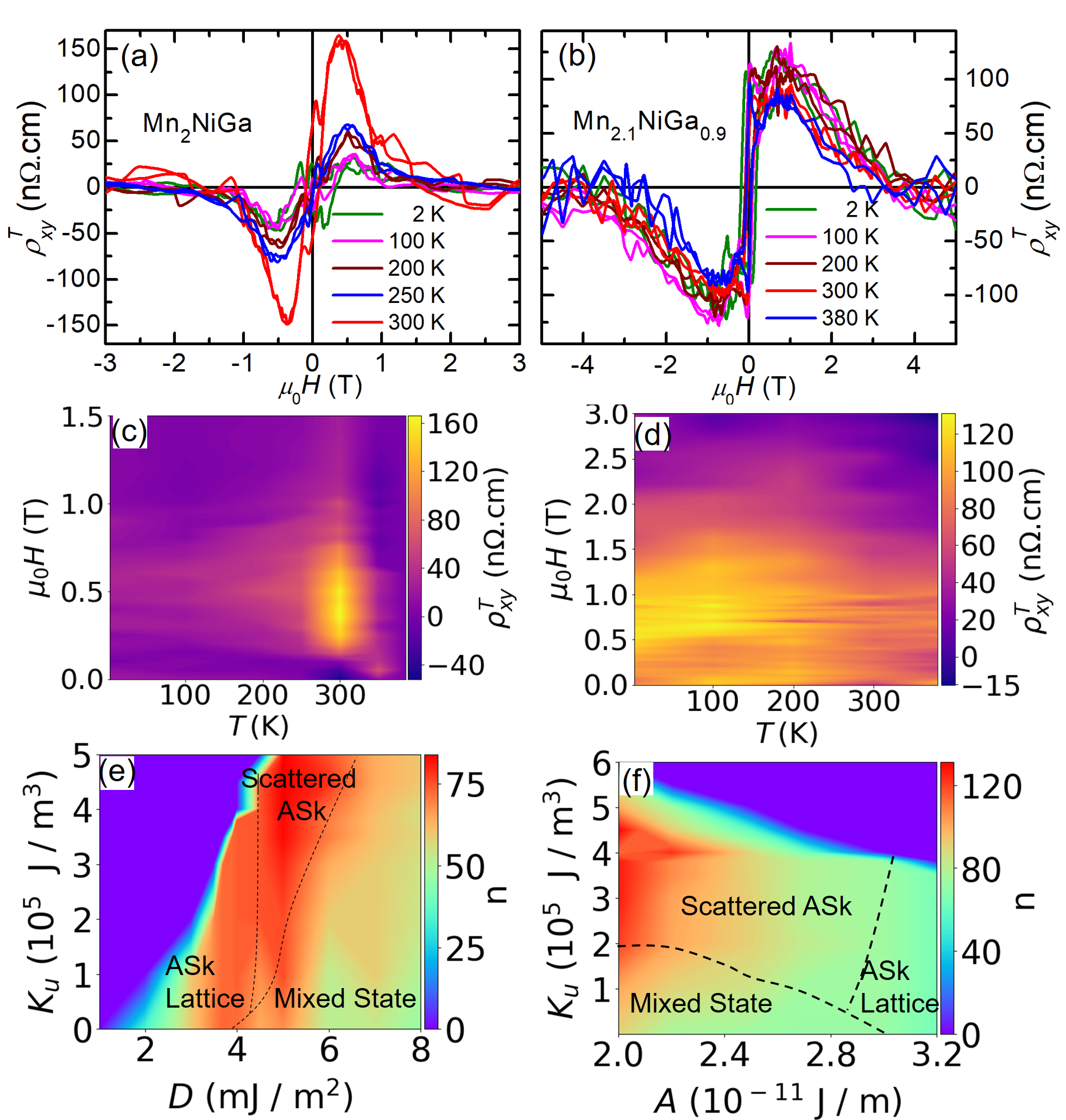}
	\caption{\label{FIG1}(Color online) Field dependent topological Hall resistivity ($\rho^T_{xy}$) at different temperatures for (a) Mn$_2$NiGa and (b) Mn$_{2.1}$NiGa$_{0.9}$. The $H-T$ phase diagram showing field dependence of  $\rho^T_{xy}$ at different temperatures derived from the topological Hall effect measurements  for (c) Mn$_2$NiGa and (d) Mn$_{2.1}$NiGa$_{0.9}$. (e) Magnetic anisotropy $K\textsubscript{u}$ versus DMI,  (f) $K\textsubscript{u}$ versus exchange stiffness constant ($A$), phase diagrams illustrating the stability of antiskyrmion (ASx) phase for different $K\textsubscript{u}$, $D$ and $A$ values. Dotted lines represent bounday between different phases.  Different colors in the phase diagram represent number density of antiskyrmions ($n$).}
\end{figure}

A factor that significantly contributes to the size and stability of the antiskyrmions in the present tetragonal materials is the anisotropy energy. The presence of considerable amount of magnetic anisotropy in the present system can be seen from the out-of-plane type hysteretic  behavior of the $M(H)$ loop in Fig. 1(d). A slight change in the Mn/Ga ratio significantly changes the coercive field and the magnetic ordering temperature. This signifies a considerable change in the magnetic anisotropy as well as the exchange constant $J$, which  can modify the size and stability of the antiskyrmion phase. For a detailed understanding of the stability of antiskyrmion phase at different anisotropy constant $K\textsubscript{u}$, DMI constant $D$ ($D_x=-D_y$), and exchange stiffness constant $A$,  we have carried out a detailed micromagnetic simulation using public domain software package Object Oriented MicroMagnetic Framework (OOMMF) \cite{oommf} with DMI extension module \cite{dmi}. Initially, a $1000 \times 1000 \times 2$~nm$^3$ thin film was relaxed from a random magnetization state in presence of perpendicular magnetic field for different values of $D$ and $A$  with zero anisotrpy. $A$ was calculated from the $T\textsubscript{C}$ and  saturation magnetization $M\textsubscript{s}$ from $M(H)$ loop at 2~K.  After the initial relaxation, the anisotropy constant $K\textsubscript{u}$ was increased to various values to check the stability of antiskyrmion lattice at the corresponding values of $D$, $A$  and $K\textsubscript{u}$.
Figure 4(e) shows the $K\textsubscript{u}$-$D$ phase diagram corresponding to the experimental parameters $A  =3.0\times10\textsuperscript{-11}$~J/m , $M\textsubscript{s} = 2.37\times10\textsuperscript{5}$~A/m, and $\mu_0H$ = 500~mT. A stable antiskyrmion lattice can be found for $D \approx 2.0-4.0$~mJ/m$^{2}$ and $K\textsubscript{u} \approx 0-5\times10\textsuperscript{5}$~J/m$^3$. A rough estimation of the anisotropy constant from the $M(H)$ loops yields $K\textsubscript{u} \approx 3.0-5.0\times10\textsuperscript{5}$~J/m$^3$. A higher $D$ and $K\textsubscript{u}$ results in a mixed  or scattered antiskyrmion phase. For a fixed $D=4.0$~mJ/m$^{2}$, a stable antiskyrmion lattice can be found for $A  \approx3.0\times10\textsuperscript{-11}$~J/m [Fig. 4(f)]. At lower values of $A$, which is expected for Mn$_{2.1}$NiGa$_{0.9}$, mixed phase and scattered antiskyrmions were stabilized at higher values of $K\textsubscript{u}$.  A decrease in the size of antiskyrmions at lower $A$ and higher $K\textsubscript{u}$ leads to a significant increase in the density ($n$) even in the mixed and scattered antiskyrmion state.

As it can be seen, a stable antiskyrmion phase can be formed for $D = 2.0-4.0$~mJ/m$^{2}$ and  $K\textsubscript{u}= 0-5\times10\textsuperscript{5}$~J/m$^3$. The size (diameter) of the antiskyrmions corresponding to these values of $D$ and $K\textsubscript{u}$ is about 40-60~nm. It is known that the magnitude of topological Hall voltage greatly depends upon the size  and density of skyrmions. We have estimated the size of the antiskyrmions from the measured topological Hall effect using the relation $\rho^T_{xy}=PR_0B_{eff}$, where $P$ is the conduction electron spin polarization and $B_{eff}$ is the effective (fictitious) magnetic field \cite{Neubauer09}. Further $B_{eff}$ can be expressed as $B_{eff}=-\phi_0/a_{sk}$, with $\phi_0=h/e$ is the magnetic flux generated by a single skyrmion and $a_{sk}$ is the size of the skyrmion \cite{Kanazawa15}. The conduction electron polarization can be roughly estimated as $P=M_{sp}/M_s$, where $M_{sp}$ is the ordered moment in the antiskyrmion phase and $M_s$ is the saturation magnetic moment in the system \cite{Neubauer09}. In the present case $P$ comes about 0.7.  By taking the highest THE at room temperature for Mn$_2$NiGa, the effective magnetic field is calculated as 8.8~T and the size of the antiskyrmions is found to be about 22~nm. A small mismatch of the antiskyrmion size  might be arising from the fact that the simulations were carried for a thin film, whereas, experiments were performed on the bulk materials. It is worth to mention here that the size of the antiskyrmions in  Mn-Ni-Ga system  is much smaller in comparison to the recently observed antiskyrmion size of 150~nm in Mn-Pt(Pd)-Sn based Heusler materials \cite{Nayak17}. In the present case, by slightly changing the composition, both the magnetic anisotropy and the exchange interaction can be tuned significantly. This is evident from the increase in the coercive field and decrease in the $T_{\text C}$ as well as the saturation magnetic moment; which in principle could give rise to a reduced antiskyrmion size in Mn$_{2.1}$NiGa$_{0.9}$. Due to the ferrimagnetic ordering in the present system one can expect a reduced skyrmion Hall effect~\cite{Woo18}, that might help the present ferrimagnetic antiskyrmions to move along the direction of applied currents. 


In summary, we have established the presence of a large topological Hall effect that withstands above room temperature in the Mn-Ni-Ga based magnetic shape memory alloys. The topological Hall effect that exists in the $D_{2d}$ symmetry based tetragonal phase vanishes when the system undergoes a structural transition to the cubic phase. The origin of the observed THE is attributed to the presence of magnetic antiskyrmions. Owing to the large out-of-plane magnetic anisotropy in the present tetragonal phase,  a detailed micromagnetic simulation was carried out to understand the stability of antiskyrmion phase in presence of different exchange interaction strength, anisotropy, and DMI. The present ferrimagnetic antiskyrmions with very high ordering temperatures possess a  great potential for their application in {\it racetrack} memory devices as they are expected to display reduced skyrmion Hall effect in comparison to the ferromagnetic ones.


\begin{acknowledgments}
	
This work was financially supported by Department of Atomic Energy
(DAE) and Department of Science and Technology (DST)-Ramanujan research grant (No. SB/S2/RJN-081/2016) of
the Government of India.

\end{acknowledgments}


\end{document}